\documentclass[]{aa}
\usepackage{graphics,latexsym,epsfig,graphicx}
\begin{document}
\title{The covariance of cosmic shear correlation functions and
  cosmological parameter estimates using redshift information}
\author{Patrick Simon, Lindsay King \& Peter Schneider}
\institute{Institut f{\"u}r Astrophysik und Extraterrestrische
  Forschung, Universit{\"a}t Bonn, Auf dem H{\"u}gel 71, D-53121 Bonn,
  Germany\\ } \date{} \authorrunning{Simon, King \& Schneider}
\titlerunning{3D correlations of cosmic shear} \abstract{ Cosmological
  weak lensing by the large scale structure of the Universe, cosmic
  shear, is coming of age as a powerful probe of the parameters
  describing the cosmological model and matter power spectrum. It
  complements Cosmic Microwave Background studies, by breaking
  degeneracies and providing a cross-check. Furthermore, upcoming
  cosmic shear surveys with photometric redshift information will
  enable the evolution of dark matter to be studied, and even a crude
  separation of sources into redshift bins leads to improved
  constraints on parameters.  An important measure of the cosmic shear
  signal are the shear correlation functions; these can be directly
  calculated from data, and compared with theoretical expectations for
  different cosmological models and matter power spectra. We present a
  Monte Carlo method to quickly simulate mock cosmic shear surveys.
  One application of this method is in the determination of the full
  covariance matrix for the correlation functions; this includes
  redshift binning and is applicable to arbitrary survey geometries.
  Terms arising from shot noise and cosmic variance (dominant on small
  and large scales respectively) are accounted for naturally. As an
  illustration of the use of such covariance matrices, we consider to
  what degree confidence regions on parameters are tightened when
  redshift binning is employed. The parameters considered are those
  commonly discussed in cosmic shear analyses - the matter density
  parameter $\Omega_{\rm m}$, dark energy density parameter (classical
  cosmological constant) $\Omega_{\Lambda}$, power spectrum
  normalisation $\sigma_{8}$ and shape parameter $\Gamma$. We
  incorporate our covariance matrices into a likelihood treatment, and
  also use the Fisher formalism to explore a larger region of
  parameter space. Parameter uncertainties can be decreased by a
  factor of $\sim 4-8$ ($\sim 5-10$) with 2 (4) redshift bins.
   } \def\A{{\cal A}}
\def\eck#1{\left\lbrack #1 \right\rbrack} \def\eckk#1{\bigl[ #1
  \bigr]} \def\rund#1{\left( #1 \right)} \def\abs#1{\left\vert #1
  \right\vert} \def\wave#1{\left\lbrace #1 \right\rbrace}
\def\ave#1{\left\langle #1 \right\rangle} \def\arcsecf
{\hbox{$.\!\!^{\prime\prime}$}} \def\arcminf {\hbox{$.\!\!^{\prime}$}}
\def\bet#1{\left\vert #1 \right\vert} \def\vp{\varphi}
\def\vt{{\vartheta}} \def\map{{$M_{\rm ap}$}} \def\d{{\rm d}}
\def\mj{$\rm {m_{j}}$} \def\mk{$\rm {m_{k}}$} \def\col{$\rm
  {m_{j}}-\rm {m_{k}}$}\def\eps{{\epsilon}} \def\vc{\vec}
\def\be{\begin{equation}} \def\ee{\end{equation}} \def\s{{\rm d}}
\def\s{{\rm s}} \def\t{{\rm t}} \def\E{{\rm E}} \def\L{{\cal L}}
\def\s8{{\sigma_{8}}} \def\om{{\Omega_{\rm m}}} \def\i{{\rm i}}
\def\seps{{\sigma_{\epsilon}}} \newcommand{\Ref}[1]{(\ref{#1})}
\newcommand{\Cite}[1]{[\cite{#1}]} \newcommand{\Vector}[1]{{\bf #1}}
\newcommand{\Matrix}[1]{{\bf #1}} {\catcode`\@=11
  \gdef\SchlangeUnter#1#2{\lower2pt\vbox{\baselineskip 0pt
      \lineskip0pt
      \ialign{$\m@th#1\hfil##\hfil$\crcr#2\crcr\sim\crcr}}}}
\def\gtrsim{\mathrel{\mathpalette\SchlangeUnter>}}
\def\lesssim{\mathrel{\mathpalette\SchlangeUnter<}} \maketitle

\section{Introduction}
The statistics of the distorted images of distant galaxies,
gravitationally lensed by the tidal gravitational field of intervening
matter inhomogeneities, contain a wealth of information about the
power spectrum of the dark and luminous matter in the Universe, and
the underlying cosmological parameters. The importance of ``cosmic
shear'' as a cosmological tool was proposed in the early 1990s by
Blandford et al. (1991), Miralda-Escud\'e (1991) and Kaiser (1992).
Further analytic and numerical work (e.g.. Kaiser 1998; Schneider et
al. 1998; White \& Hu 2000) took into account the increased power on
small scales, resulting from the non-linear evolution of the power
spectrum (Hamilton et al. 1991; Peacock \& Dodds 1996).

The feasibility of cosmic shear studies was demonstrated in 2000, when
four teams announced the first observational detections (Bacon et al.
2000; Kaiser et al. 2000; van Waerbeke et al. 2000; Wittman et al.
2000). Upcoming surveys will cover much larger areas, and multicolour
observations will enable photometric redshift estimates for the
galaxies to be obtained. For example, the CFHT Legacy survey
(http://www.cfht.Hawaii.edu/Science/CFHLS) will cover 172 deg$^2$ in 5
optical bands, with a smaller area to be observed in J and K bands.

In order to compare these observations with predictions for various
cosmological models and matter power spectra, different two-point
statistics of galaxy ellipticities can be employed, all of which are
filtered versions of the convergence power spectrum. Here, we focus on
the gravitational shear correlation functions, which can be directly
obtained from the data as described in Sect.\,\ref{scf}.

This quest for the parameters describing the matter content and
geometry of the Universe is limited by several sources of error,
dominated by the dispersion in the intrinsic ellipticities of galaxies
and by cosmic (sampling) variance.  The covariance (error) matrix is
thus an essential ingredient in the extraction of parameters from
data, or in parameter error estimate predictions. Schneider et al.
(2002a) provide analytical approximations for the contributions to the
covariance matrix. They consider the case when the mean redshift of
the population is known, and data taken in a single contiguous area.
Kilbinger \& Schneider (2004) use a numerical approach to investigate
the impact of survey geometry on parameter constraint. Using a Fisher
matrix approach, which provides a lower-bound estimate of covariance,
Hu (1999) has shown that even crude redshift information enables much
tighter constraints to be placed on cosmological parameters, compared
with the case when only the mean redshift of the population is known.
This study concentrated on the convergence power spectrum as the
vehicle of cosmological information.

Motivated by these studies, in this paper we demonstrate how numerical
simulations can be used to estimate the full covariance matrix for the
shear correlation functions in the presence of redshift information,
and for arbitrary survey geometries.  We consider auto- and
cross-correlations for redshift bins (as in Hu 1999) and in addition
allow for cross-correlations between measurements of the shear signal
at different angular scales. With covariance matrices in hand, we then
investigate the improvement in parameter estimates due to redshift
binning.
 
Further details and derivations of the equations relevant to cosmic
shear and weak lensing can be found in Bartelmann \& Schneider (2001).
For a recent review of cosmic shear and future prospects see van
Waerbeke \& Mellier (2003).

\section{Power spectrum and correlation functions}
Access to cosmological parameters is provided through the {\em
  observable} two-point statistics of the ellipticities of distant
galaxies. In this section, we describe how these are related to the
matter power spectrum, and to the underlying density field.

\subsection{The convergence power spectrum}

The power spectrum $P_{\kappa}(\ell)$ of the effective
convergence, or equivalently of the shear $P_{\gamma}(\ell)$ (e.g.
Bartelmann \& Schneider 2001), is related to that of the density
fluctuations, $P_{\delta}(\ell)$, through a variant of Limber's
equation in Fourier space (Kaiser 1998)
\begin{eqnarray}
  \label{kappow}
  P_\kappa(\ell)&=&
  \frac{9H_0^4\Omega_{\rm m}^2}{4c^4}\, \int_0^{w_\mathrm{H}}\,\d
  w\,\frac{\bar{W}^2(w)}{a^2(w)}\,
  P_\delta\left(\frac{\ell}{f(w)},w\right)
  \\
  \label{bigw}
  \bar{W}(w)&\equiv&\int_w^{w_\mathrm{H}}\d w'\,p(w')\,
  \frac{f(w'-w)}{f(w')} 
  \\
  f(w) &=& \left\{
  \begin{array}{ll}
    K^{-1/2}\sin(K^{1/2}w) & (K>0)\\
    w & (K=0)\\
    (-K)^{-1/2}\sinh[(-K)^{1/2}w] & (K<0) \\
  \end{array}\right.
\; ,
\end{eqnarray}
where $\ell$ is the angular wave-vector, Fourier space conjugate to
$\vec\theta$. $w$ is the comoving radial distance, $K$ the curvature
parameter. A value $K=0$ corresponds to $\Omega_{\rm
  m}\;+\;\Omega_{\Lambda}=1$, where $\Omega_{\Lambda}$ is the
cosmological constant and $\Omega_{\rm m}$ the matter density
parameter. The function $\bar{W}(w)$ accounts for the sources being
distributed in redshift, where $p(w')dw'$ is the comoving distance
probability distribution for the sources. $a\left(w\right)$ is the
scale factor normalised to $a\left(w=0\right)=1$ and $H_0$ is the
Hubble constant.

Splitting up the weak lensing survey in redshift, as in
Fig.\ref{redshiftbins}, defines a set of effective convergence and
shear maps instead of a single one, giving more information on the
evolution of the dark matter fluctuations and therefore enabling
tighter constraints to be placed on cosmological parameters.  Auto-
and cross-correlation of these maps introduce a whole set of power
spectra, generalising Eq.\,\Ref{kappow}:
 \begin{eqnarray}\label{powerspectra}
  P^{\left(ij\right)}_\kappa\left(\ell\right)&=&
  \frac{9H_0^4\Omega_{\rm m}^{2}}{4c^4}\nonumber
  \\\nonumber
  &\times&\int_0^{w_H} \d w\,
  \frac{\bar{W}^{\left(i\right)}\left(w\right)\bar{W}^{\left(j\right)}\left(w\right)}
  {a^2\left(w\right)}
  P_\delta\left(\frac{\Vector{\ell}}{f\left(w\right)},w\right)
  \\
  \bar{W}^{\left(i\right)}\left(w\right)&\equiv&
  \int_{w_{i-1}}^{w_i}\d w^\prime~p^{\left(i\right)}\left(w^\prime\right)
  \frac{f\left(w-w^\prime\right)}{f\left(w^\prime\right)}
  \;,
\end{eqnarray}
with $p^{\left(i\right)}\left(w\right)$ being the normalised
distribution in comoving distance inside the $i$th bin, where $i$ runs
between $1$ and the number of redshift bins $N_{z}$.
$P_\kappa^{\left(ii\right)}$ are auto-correlation power spectra,
whereas $P_\kappa^{\left(ij\right)}$ with $i\ne j$ are
cross-correlation power spectra.

\subsection{Shear correlation functions\label{scf}}
Constraints can be placed on cosmological parameters using the {\em
  directly observable} shear correlation functions, which we now turn
to.

The basis which underpins the use of the distorted images of
distant galaxies in weak lensing studies is a transformation
relating the source, $\eps^{\rm (s)}$, and image, $\eps$, (complex)
ellipticities to the tidal gravitational field of density
inhomogeneities (for definitions see Bartelmann~\&~Schneider 2001).
We focus on the non-critical regime where
\begin{equation}
  \eps={\eps^{\rm (s)} + g\over 1+g^{*}\eps^{\rm (s)}}\approx \eps^{\rm (s)} + \gamma\;,
  \label{eps}
\end{equation}
where $g\equiv\gamma/(1-\kappa)$ is the reduced shear.

Empirically the probability distribution function (pdf) of the
galaxies' intrinsic ellipticities is a truncated Gaussian for both the
real and imaginary parts of $\epsilon^{\rm (s)}$:
\begin{equation}\label{intrinsicpdf}
p_{\epsilon^{\rm (s)}}=
\frac{\exp\rund{-|\eps^{\rm (s)}|^{2}/\sigma_{\eps^{\rm (s)}}^{2}}}
{\pi\sigma^{2}_{\eps^{\rm (s)}}\left[1-\exp\rund{-1/\sigma_{\eps^{\rm
          (s)}}^2}\right]}
\;,
\end{equation}
where $\sigma_{\eps^{\rm (s)}}$ is the intrinsic ellipticity
dispersion.

As in Schneider et al. (2002b), the shear correlation
functions are defined as \be \xi_\pm(\theta)=\ave{\gamma_{\rm
    t}\gamma_{\rm t}}\pm\ave{\gamma_\times
  \gamma_\times}=\int_0^\infty {\d\ell\,\ell\over 2\pi}\,{\rm
  J}_{0,4}(\ell\theta)\;P_{\kappa}(\ell)\;,
\label{corre}\\
\ee where J$_{n}$ are $n$-th order Bessel functions of the first kind;
$\gamma_{\rm t}$ and $\gamma_\times$ are the tangential and cross
shear components respectively.  From now on we focus on $\xi_{+}$,
since this contains most of the cosmological information on the scales
of interest.

\subsection{Choice of cosmology and matter power spectrum}

Unless otherwise stated, our cosmology throughout is a $\Lambda$CDM
model with $\Omega_{\rm m}=0.3$, $\Omega_{\Lambda}=0.7$ and
$H_{0}=70\,{\rm km}\,{\rm s}^{-1}\,{\rm Mpc}^{-1}$.  A scale-invariant
($n=1$, Harrison-Zel'dovich) spectrum of primordial fluctuations is
assumed. Predicting the shear correlation functions requires a model
for the redshift evolution of the 3-D power spectrum.  We use the
fitting formula of Bardeen et al. (1986; BBKS) for the transfer
function, and the Peacock and Dodds (1996) prescription for evolution
in the nonlinear regime. The power spectrum normalisation is
parameterised with $\sigma_{8}=0.9$, and $\Gamma=0.21$. Quantities
calculated for this fiducial cosmology/power spectrum will be
super-scripted with a ``t".

\section{Simulating cosmic shear surveys}

In this section we describe the method we used to make Monte-Carlo
simulations of cosmic shear surveys. The implementation makes the
simulations as computationally inexpensive as possible, i.e. without
invoking N-body simulations.

Calculating the lensing signal by ray tracing through N-body
simulations has become a common tool for making simulated weak lensing
surveys (see e.g. Blandford et al. 1991; Wambsganss, Cen~\&~Ostriker
1998; Jain et al. 2000).

We take a different path here, because only the two-point statistics
of weak lensing is considered. This allows us to reduce the
computational effort by expressing the fields of the shear and
convergence as random Gaussian fields having the same power spectrum
as the corresponding fields from the N-body approach. In the weak
lensing regime, ray-tracing is well described by the Born
approximation which ignores the effects of lens-lens coupling and
deviations of light rays from the fiducial path (see White~\&~Hu
2000). The task of calculating the required power spectra then becomes
relatively straightforward, because these can be shown to be linear
functions of the three-dimensional evolving dark matter power
spectrum. The accuracy of the results depends on how accurately that
three-dimensional power spectrum is known.

Since we consider the two-point cosmic shear statistics in an area of
relatively small angular size, we can represent the cosmic shear
fields by random Gaussian fields in a flat sky approximation.

The practical advantage of simulating a single random Gaussian field
$\delta\left(\Vector{r}\right)$ - a \emph{homogeneous} and isotropic
random Gaussian to be exact - is the fact that from the Fourier
coefficients
\begin{equation}\label{coeff}
  c_\Vector{k}=\frac{1}{V}\int_V
  \d\Vector{r}~
  \delta\left(\Vector{r}\right)\exp{\left({\rm i}\Vector{k}\cdot\Vector{r}\right)}
\end{equation}
of such a random field only a pair of coefficients is correlated
\begin{equation}\label{cond1}
\langle c_\Vector{k}c_{-\Vector{k}^\prime}\rangle=
\frac{1}{V}\delta_{\Vector{k}\Vector{k}^\prime}
P_\Vector{k}
\;,
\end{equation}
where $\delta_D$ is the Dirac delta function and $P_\Vector{k}$ is the
\emph{power spectrum} of the random field. The volume $V$ of the
Gaussian field is in our case simply the area on the sky covered by
the field.

In our work, however, the situation is a bit more complicated than
that: the ellipticities of galaxies belonging to different redshift
bins are correlated as well as the ellipticities of galaxies at
different angular positions. Therefore, when more than one redshift
bin is considered, several Gaussian fields - cosmic shear maps - with
prescribed cross-correlations have to be simulated
\emph{simultaneously}. A way to do this in general on a regular grid
for real Gaussian fields is shown in the following subsection.  The
subsection thereafter explains how we used this approach for
simulating mock cosmic shear surveys.

\subsection{Realisations of Correlated Gaussian Fields}

According to condition \Ref{cond1} the pair $c_\Vector{k}$ and
$c_\Vector{-k}$ is correlated. In this section we restrict ourselves
to real Gaussian fields with
$\delta\left(\Vector{r}\right)=\delta^\ast\left(\Vector{r}\right)$.
This introduces an additional condition that follows from the
definition \Ref{coeff} of the $c_\Vector{k}$:
\begin{equation}\label{cond2}
  c^{}_\Vector{k}=c^\ast_\Vector{-k}
\; .
\end{equation}
In particular, for real Gaussian fields we thereby have
\begin{equation}
  \langle c_\Vector{k}^{}c^\ast_{\Vector{k}^\prime}\rangle=
  \frac{1}{V}P_\Vector{k}\delta_{\Vector{k}\Vector{k}^\prime}  
  \;,
\end{equation}
$\delta_{\Vector{k}\Vector{k}^\prime}$ being the Kronecker symbol.
The conditions \Ref{cond1} and \Ref{cond2} are easily accounted for
if, say, only the $c_\Vector{k}$ for half of the spatial frequencies
$\Vector{k}$ are worked out and the $c_\Vector{-k}$ frequencies are
set accordingly. Hence, for our choice, if we talk about
$c_\Vector{k}$ we actually mean only Fourier coefficients in the right
half-plane.

Furthermore, the real and imaginary parts of $c_\Vector{k}$ are
uncorrelated, and both follow the same Gaussian pdf.  This pdf has
zero mean\footnote{In the case that $\delta\left(\Vector{r}\right)$
  has a non zero mean, $\langle c_\Vector{k}\rangle$ for
  $\Vector{k}=0$ becomes different from zero.}  and a variance
$\sigma_\Vector{k}$ that is expressed in terms of the power spectrum
$P_\Vector{k}$ describing the two-point correlations of the
fluctuations in the Gaussian field (see e.g. Peacock 2001)
\begin{equation}
  \sigma^2_\Vector{k} = \frac{1}{2V}P_\Vector{k}
  \; .
\end{equation}

The procedure for making \emph{one} Gaussian field realisation
requires two steps: 1. drawing numbers for the real and imaginary
parts for every $c_\Vector{k}$ with a Gaussian random number
generator, and 2.  transformation of this Fourier space representation
to real space in order to obtain the field realisation. For the second
step we used an FFT algorithm from Press et al. (1992)\footnote{ As in
  FFT the matrix of the Fourier coefficients contains $c_\Vector{k}$
  that share the same matrix elements with $c_\Vector{-k}$, one has
  for these particular coefficients to set the imaginary parts to zero
  and to increase $\sigma_\Vector{k}$ by the factor $\sqrt{2}$. The
  latter is necessary to guarantee that the variance of the modulus of
  $c_\Vector{k}$ is still correct.}.

This procedure also holds when realisations of more than one, but
\emph{uncorrelated} Gaussian fields are desired. ``Uncorrelated''
means that if we denote the Fourier coefficients of, say, $N$ Gaussian
random fields by $c^{(i)}_\Vector{k}$ with $i=1..N$ then we expect for
those fields the relation
\begin{equation}
\Big\langle 
c^{\left(i\right)}_\Vector{k}\left[c^{\left(j\right)}_{\Vector{k}^\prime}\right]^\ast\Big\rangle=
\frac{1}{V}P^{\left(ii\right)}_\Vector{k}\delta_{ij}
\delta_{\Vector{k}\Vector{k}^\prime}  
\;,
\end{equation}
where $\delta_{ij}$ is also a Kronecker symbol, this time for the
Gaussian field indices. $P^{\left(ii\right)}_\Vector{k}$ is the
previously introduced power spectrum, or \emph{auto-correlation} power
spectrum, of the $i$th random field. Thus, here correlations between
$c_\Vector{k}$ of different random fields vanish.

For the purposes of this work, however, we need to be able to allow
for cross-correlations between different random fields $i\neq j$ in a
defined manner, like
\begin{equation}\label{eqn1}
\Big\langle 
c^{\left(i\right)}_\Vector{k}\left[c^{\left(j\right)}_{\Vector{k}^\prime}\right]^\ast\Big\rangle=
\frac{1}{V}P^{\left(ij\right)}_\Vector{k}
\delta_{\Vector{k}\Vector{k}^\prime}  
\; .
\end{equation}
$P^{\left(ij\right)}_\Vector{k}$ is for $i\neq j$ the
\emph{cross-correlation} power spectrum. Like for the
auto-correlations, only certain pairs of Fourier coefficients of
different Gaussian fields are correlated. This follows from the
assumption that the cross-correlations are homogeneous, too. Note that
$P^{\left(ij\right)}_\Vector{k}=P^{\left(ji\right)}_\Vector{k}$.

In order to find a recipe for making realisations of that kind, we
make the Ansatz that the $N$ Fourier coefficients
$c^{\left(i\right)}_\Vector{k}$ are a \emph{linear transformation}
$\Matrix{A}_\Vector{k}$ of $N$ different uncorrelated coefficients
$d^{\left(i\right)}_\Vector{k}$ with an equal Gaussian pdf for the
real and imaginary parts, zero mean and a $1/\sqrt{2}$ dispersion
\begin{equation}\label{eqn2}
  \Big\langle 
  d^{\left(i\right)}_\Vector{k}
  \left[d^{\left(j\right)}_\Vector{k}\right]^\ast
  \Big\rangle=\delta_{ij}\; ;\;\;
  c^{\left(i\right)}_\Vector{k}=
  \sum_q\left[\Matrix{A}_\Vector{k}\right]^i_{~q}~d^{\left(q\right)}_\Vector{k}
\; .
\end{equation}
$\Matrix{A}_\Vector{k}$ is a $N\times N$ linear transformation matrix.
The linearity of this transformation accounts for the fact that the
resulting set of coefficients $c^{\left(i\right)}_\Vector{k}$ still
obeys a Gaussian statistics, because linear combinations of Gaussian
random variables are also Gaussian.

Since for real Gaussian fields real and imaginary parts of the Fourier
coefficients are not correlated, and by our Ansatz neither are the
real and imaginary parts of $d^{\left(i\right)}_\Vector{k}$, only real
numbers for the components $\left[\Matrix{A}_\Vector{k}\right]^i_{~q}$
are allowed; an additional imaginary part of $\Matrix{A}_\Vector{k}$
would mix real and imaginary parts of $d^{\left(i\right)}_\Vector{k}$,
thereby possibly introducing correlations between real and imaginary
parts in $c_\Vector{k}$. A matrix $\Matrix{A}_\Vector{k}$ that is
purely imaginary would be an alternative choice, though.

Eqs.\,\Ref{eqn2} can now be combined to give
\begin{eqnarray}\nonumber
 \Big\langle
 c^{\left(i\right)}_\Vector{k}\left[c^{\left(j\right)}_\Vector{k}\right]^\ast
 \Big\rangle
 &=&
 \sum_{q,r}
 \Big\langle
 \left[\Matrix{A}_\Vector{k}\right]^i_{~q}d^{\left(q\right)}_\Vector{k}
 \left[\Matrix{A}_\Vector{k}\right]^j_{~r}\left[d^{\left(r\right)}_\Vector{k}\right]^\ast
 \Big\rangle 
\\\nonumber
&=&
\sum_{q,r}
\left[\Matrix{A}_\Vector{k}\right]^i_{~q}
\left[\Matrix{A}_\Vector{k}\right]^j_{~r}
\Big\langle
d^{\left(q\right)}_\Vector{k}
\left[d^{\left(r\right)}_\Vector{k}\right]^\ast
\Big\rangle
\\\nonumber
&=&
\sum_{q,r}
\left[\Matrix{A}_\Vector{k}\right]^i_{~q}
\left[\Matrix{A}_\Vector{k}\right]^j_{~r}\delta_{qr}
\\
&=&
\sum_{q}
\left[\Matrix{A}_\Vector{k}\right]^i_{~q}
\left[\Matrix{A}_\Vector{k}^T\right]^q_{~j}
\; ,
\end{eqnarray}
where $\Matrix{A}_\Vector{k}^T$ denotes the transpose of
$\Matrix{A}_\Vector{k}$.  Hence, together with equation \Ref{eqn1}
this puts another constraint on the matrix $\Matrix{A}_\Vector{k}$,
namely
\begin{equation}
  \frac{1}{V}P^{\left(ij\right)}_\Vector{k}=
  \sum_q
  \left[\Matrix{A}_\Vector{k}\right]^i_{~q}
  \left[\Matrix{A}_\Vector{k}^T\right]^q_{~j}\;.
\end{equation}
For convenience we introduce the \emph{power matrix} defined as
$\left[\Matrix{P}_\Vector{k}\right]^i_{~j}\equiv
\frac{1}{V}P^{\left(ij\right)}_\Vector{k}$ to abbreviate this
equation:
\begin{equation}\label{eqn3}
  \Matrix{P}_\Vector{k}=\Matrix{A}_\Vector{k}\Matrix{A}_\Vector{k}^T\;.
\end{equation}
The power matrix is the covariance matrix between the Fourier
coefficients of a set of Gaussian fields for a certain $\Vector{k}$.

This shorthand of $N^2$ equations does not uniquely determine the
matrix $\Matrix{A}_\Vector{k}$, because it contains only $N(N+1)/2$
linearly independent equations, since both the matrix on the lhs and
the matrix product on the rhs are symmetric.  As there are no further
constraints on $\Matrix{A}_\Vector{k}$, we are allowed to set the
remaining $N^2-N(N+1)/2=N(N-1)/2$ constraints of
$\Matrix{A}_\Vector{k}$ as we like. We do this by assuming that
$\Matrix{A}_\Vector{k}$ is symmetric, so that we finally obtain
\begin{equation}\label{eqn4}
  \Matrix{P}_\Vector{k}=
  \Matrix{A}_\Vector{k}^2~~\Rightarrow~~
  \Matrix{A}_\Vector{k}=\sqrt{\Matrix{P}_\Vector{k}}
\; .
\end{equation}

In general the square root is not unique (see e.g. Higham 1997).
However, we are already satisfied with one particular solution to this
problem. In order to determine such a solution, note that
$\Matrix{P}_\Vector{k}$ is a symmetric positive (semi)definite matrix,
which is ensured by the properties of the power spectra the power
matrix consists of:
\begin{eqnarray}
  P^{(ij)}_\Vector{k}&=&P^{(ji)}_\Vector{k}\;,\\
  \left[P^{(ij)}_\Vector{k}\right]^2&\le&
  P^{(ii)}_\Vector{k}P^{(jj)}_\Vector{k}
  \; .
\end{eqnarray}
Therefore, this matrix can uniquely be decomposed into
\begin{equation}
  \Matrix{P}_\Vector{k}=
  \Matrix{R}_\Vector{k}^T~\Matrix{D}_\Vector{k}~\Matrix{R}_\Vector{k}
  \; ,
\end{equation}
where $\Matrix{R}_\Vector{k}$ is an orthogonal matrix whose column
vectors are the eigenvectors of $\Matrix{P}_\Vector{k}$, while their
corresponding, always real and positive, eigenvalues $\lambda_i$ are
on the diagonal of the diagonal matrix $\Matrix{D}_\Vector{k}=\rm
diag\left(\lambda_1,\lambda_2,...,\lambda_N\right)$. As one particular
square root we pick out
 \begin{eqnarray}\nonumber
   \Matrix{A}_\Vector{k}&=&
   \Matrix{R}_\Vector{k}^T~\sqrt{\Matrix{D}_\Vector{k}}~\Matrix{R}_\Vector{k}
   \\
   \sqrt{\Matrix{D}_\Vector{k}}&\equiv&{\rm
   diag}\left(\sqrt{\lambda_1},\sqrt{\lambda_2},...,\sqrt{\lambda_N}\right)
\end{eqnarray}
which is a solution due to
$\sqrt{\Matrix{D}_\Vector{k}}\sqrt{\Matrix{D}_\Vector{k}}=
\Matrix{D}_\Vector{k}$.

To sum up, for every $\Vector{k}$ mode considered, the process for the
realisation of correlated Gaussian random fields requires one to find
the square root $\Matrix{A}_\Vector{k}$ of the power matrix
$\Matrix{P}_\Vector{k}$. This defines a linear transformation for a
vector of uncorrelated random complex numbers (real and imaginary part
of the same coefficient are uncorrelated, too) with zero mean, real
and imaginary parts obeying a Gaussian pdf with $1/\sqrt{2}$ variance.
Applying $\Matrix{A}_\Vector{k}$ yields a vector of Fourier
coefficients belonging to the realisations of the correlated Gaussian
random fields. Due to \Ref{cond2} this is performed only for one half
of the spatial frequencies considered. The other half is set
accordingly to fulfil this condition.

\subsection{Simulating the weak lensing survey}

\begin{center}
\begin{figure}[width=75mm]
  \epsfig{file=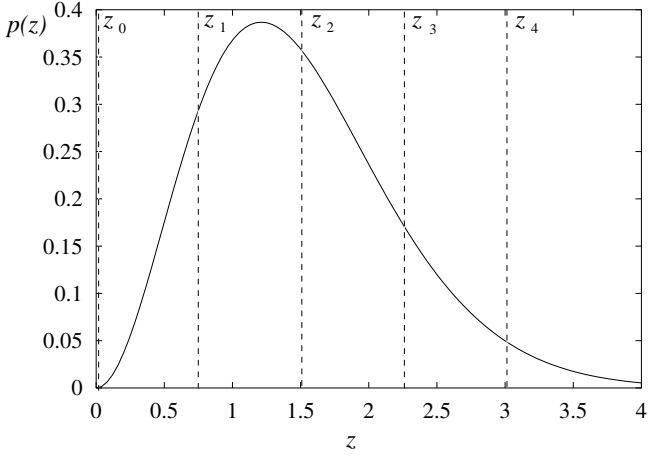,width=90mm}
\caption{Galaxies are binned together according to their redshift, 
  the boundaries of the pairwise adjacent redshift bins are $w_i$ with
  $i=0...N_{z}$ (here as an example $N_{z}=4$).  For every redshift
  bin the reduced shear field is calculated, averaging over the
  redshift distribution inside the bin.}
\label{redshiftbins}
\end{figure}
\end{center}
Each galaxy in the mock galaxy catalogue is defined by an angular
position, an ellipticity $\epsilon$ and a redshift bin it belongs to.
The ellipticity of the isophotes of a galaxy is determined by the
intrinsic shape of a galaxy $\epsilon^{\rm (s)}_i$ and the reduced
shear $g$ at the position of the galaxy (see Eq.\,\ref{eps}). The
reduced shear is a function of the convergence $\kappa$ and shear
$\gamma$ which have to be simulated for each redshift bin as a map
covering the simulated area.  Here we assume that the galaxies are
binned into $N_{z}$ pairwise adjacent redshift bins, chopping off the
redshift distribution
\begin{equation}
p\left(z\right)=\frac{1}{z_0^3}
\frac{1}{\Gamma\left(3/\beta\right)\beta}z^2
\exp{\left[-\left(\frac{z}{z_0}\right)^\beta\right]}
\label{redz}
\end{equation}
as in Fig.\,\ref{redshiftbins}.  This empirical distribution with
$\beta=1.5$ and $z_0=1.0$ is based on deep field surveys (see e.g.
Smail et al. 1995). The total number of galaxies inside the field,
with chosen size of $5^{\circ}\times5^{\circ}$, is set to be
$\approx2.7\times10^6$, to get an average of 30 galaxies per $\rm
arcmin^2$. Moreover, the galaxies are assumed to be randomly
distributed over the field of view.

The method of the last subsection is used to work out the convergence
maps in Fourier space on a grid of $2048\times2048$ pixels for all
redshift bins. As input the power matrix $\Vector{P}_\Vector{k}$,
consisting of the auto- and cross-correlation power spectra of these
convergence maps, specified by the Eqs.\,\Ref{powerspectra}, is
needed.

In the next step, the shear maps are obtained from the convergence
maps using the relation
\begin{equation}
  \tilde{\gamma}_{\vec\ell}=
  \frac{\ell_1^2-\ell_2^2+2{\rm i}\ell_1\ell_2}
  {\ell_1^2+\ell_2^2}~\tilde{\kappa}_{\vec\ell}
  ~~~~~~~
  {\vec\ell}\equiv
  \left(
    \begin{array}{l}
      {\ell_1}\\{\ell_2}
    \end{array}
  \right)
  \; .
\end{equation}
$\tilde{\gamma}_{\vec\ell}$ and $\tilde{\kappa}_{\vec\ell}$ are the
Fourier coefficients of the shear and convergence fields for the
angular frequency ${\vec\ell}$, respectively. This relation stems from
the fact that both shear and convergence are linearly related to a
potential function. For every galaxy, shear and convergence are then
combined with the intrinsic ellipticity $\epsilon^{\rm (s)}_i$,
randomly drawn from the pdf Eq.\,\Ref{intrinsicpdf} using
$\sigma_\epsilon^{\rm (s)}=0.3$, to compute the final ellipticity of
the galaxy via Eq.\,\Ref{eps}.

Both angular size $\Delta$ and number of pixels $N_P$ along one axis -
the sampling size - limits the number of fluctuation modes accounted
for in the simulated data. This means, since we are lacking
fluctuations on scales outside of
$5^\circ/\left(2N_P\right)\le\Theta\le5^\circ$, equivalent to 
$\ell_{\rm min}\le\ell\le\ell_{\rm max}$, that we have less
correlation in the cosmic shear fields than expected
(Eq.\,\ref{corre})
\begin{eqnarray}
  \xi_\pm(\theta)&=&
  \int_{\ell_{\rm min}}^{\ell_{\rm max}}
  {\d\ell\,\ell\over 2\pi}\,{\rm
    J}_{0,4}(\ell\theta)\;P_{\kappa}(\ell)
  \\
  \ell_{\rm min}&=&
  \frac{\pi}{\sqrt{2}\Delta} \; ; \; \;
  \ell_{\rm max}=
  N_P\,\ell_{\rm min}
  \; .
\end{eqnarray}
The values for the limits are estimates for a square field; the limits
are not clearly defined, because the number of $\ell$-modes in the FFT
matrix becomes very small near the cutoffs. One solution to this
problem is to artificially set a clearly defined range within the
interval $\left[\ell_{\rm min},\ell_{\rm max}\right]$, or, as we have
done, to find a best fit cutoff. This is found by varying the cutoffs
to obtain closest agreement between the theoretical two-point
correlation and the ensemble average of all Monte-Carlo realisations.

In total, we simulated two data sets. The first data set consists of
$N_f=795$ independent realisations each $5^\circ\times 5^\circ$.  The
redshift distribution of the galaxies was split into 2 bins at a
redshift cut $z_{\rm cut}=1.25$, and the distribution is truncated at
$z=3$.  For our fiducial surveys, we randomly selected $10$
sub-fields, each of $1.25^{\circ}\times1.25^{\circ}$, from different
large realisations.  This was done for two reasons: 1. to reduce the
computation time, since for $10$ shear maps we require only one
realisation, and 2. sub-fields are less affected by
$\Vector{\ell_{\min}}$ that necessarily enters the simulations due to
the finite realisation area.

The second data set has 4 redshift bins, with $z_{\rm cut}=0.75, 1.5,
2.25, 3.0$ where the last value is the truncation redshift. It has
$N_f=266$ independent realisations. The fiducial surveys from this
data set consist of single sub-fields of size
$1.25^{\circ}\times1.25^{\circ}$.

For both data sets, $\xi_+$ was estimated (see next section) for
$N_{\Delta\theta}=65$ angular separation bins, ranging from about
$2\arcminf0$ to $40\arcminf0$. For the first (second) data set, the
correlation functions were subsequently averaged for 10 (1) sub-fields
in order to simulate cosmic shear surveys consisting of 10 (1)
independent data fields.

In a further step, the cross- and auto-correlation of the cosmic shear
between the shear maps were, according to appendix \ref{xireduction},
combined to yield the cosmic shear correlations for a coarser redshift
binning; in each step the number of redshift bins was reduced by one
by combining two neighbouring bins. This process gave for the first
data set, apart from the original data, the shear correlation of one
redshift bin with boundaries $z=0$ and $z=3$.  The second data set
allows more freedom of choice for combining redshift bins, so that we
are able to construct several data sets with three and two redshift
bins. Table\,\ref{table} lists the different redshift bins and
reference names, all extracted from the two original data sets.

\begin{table}
\begin{center}
\caption{The final column denotes the name given to a particular
  binning of data. The entries in the columns $z_{i}$ show the
  corresponding cuts in redshift, $z_{\rm cut}$.\label{table}}
\begin{tabular}{lllll|l}
 $z_0$ & $z_1$ & $z_2$ & $z_3$ & $z_4$ & Name
  \\
  \hline
  &&&&&
  \\
  0 & 0.75 & 1.5 & 2.25 & 3.0 & 4bins \\
  0 & 1.5 & 2.25 & 3.0 & &  3binsI   \\
  0 & 0.75 & 2.25 & 3.0 & & 3binsII   \\
  0 & 0.75 & 1.5 & 3.0 & &  3binsIII  \\
  0 & 2.25 & 3.0 & & &      2binsI     \\
  0 & 1.5  & 3.0 & & &      2binsII     \\
  0 & 0.75 & 3.0 & & &      2binsIII    \\
  0 & 1.25 & 3.0 & & &      2binsIV   \\
  0 & 3.0 & & & &           onebin     
\end{tabular}
\end{center}
\end{table}
Below, this data is used to study the improvement in the statistical
uncertainties of the cosmological parameter estimates, if one has more
information on the redshifts of the galaxies.

\section{Estimating $\xi_+$}

To estimate the two-point correlator $\xi_+$ between the galaxy
ellipticities $\epsilon_i$ - depending on position $\Vector{\Theta}_i$
and redshift bin - inside of the sub-fields we use the estimator
\begin{eqnarray}\nonumber
  \hat{\xi}_+\left(\theta\right)
  &=&\frac{1}{N_p\left(\theta\right)}
  \\\nonumber
  &&
  \times\sum_{ij}w_iw_j
  \left(\epsilon_{i{\rm t}}\epsilon_{j{\rm t}}+
    \epsilon_{i\times}\epsilon_{j\times}\right)
  \Delta_\theta\left(\left|\Vector{\Theta}_i-\Vector{\Theta}_j\right|\right)
  \\\nonumber
  N_p\left(\theta\right)&=&
  \sum_{ij}w_iw_j
  \Delta_\theta\left(\left|\Vector{\Theta}_i-\Vector{\Theta}_j\right|\right)
  \\
  \Delta_\theta\left(\phi\right)
  &\equiv&\left\{
      \begin{array}{lr}
        1 & {\rm for~}
        \theta-\Delta\theta/2<\phi\le\theta+\Delta\theta/2 \\
        0 & {\rm otherwise}
      \end{array}
      \right.
 \end{eqnarray}
 as mentioned in Schneider et al. (2002a), with $\Delta\theta$ being
 the width of the angular bins. Since we are dealing with simulated
 data here, there is no need to weight galaxies with respect to their
 ellipticity. Therefore, we set $w_i=1$ for every galaxy.
 
 Although mathematically simple, it takes quite a time to evaluate the
 estimator due to the large number of galaxy pairs.
 To speed up the whole procedure we put a grid of
 rectangular cells of size $\Delta\theta\times\Delta\theta$ over the
 sub-field in question and compute the number $N_{ij}$ of galaxies and
 the mean of their ellipticities $\bar{\epsilon}_{ij}$ inside every
 cell. The index $ij$ indicates the position of the cell inside the
 grid. This means we are representing galaxies inside the same cell by
 a single data point with weight $N_{ij}$ and ellipticity
 $\bar{\epsilon}_{ij}$. In particular, all galaxies inside this cell
 are assumed to be placed at the same position. The estimator of
 $\xi_+$ for this rearranged data set can be shown to be
\begin{eqnarray}\nonumber
  \hat{\xi}_+\left(\theta\right)
  \!&=&\!\frac{1}{N_p\left(\theta\right)}
  \\\nonumber
  \!&\times&\!\sum_{ij,kl} 
  N_{ij}N_{kl}
  \left(\bar{\epsilon}_{ij{\rm t}}\bar{\epsilon}_{kl{\rm t}}+
  \bar{\epsilon}_{ij\times}\bar{\epsilon}_{kl\times}\right)
  \Delta_\theta\left(\left|\Vector{\Theta}_{ij}-\Vector{\Theta}_{kl}\right|\right)
  \\
  N_p\left(\theta\right)\!&=&\!
  \sum_{ij,kl}
  N_{ij}N_{kl}
  \Delta_\theta\left(\left|\Vector{\Theta}_{ij}-\Vector{\Theta}_{kl}\right|\right)
  \;,
\end{eqnarray}
where $\Vector{\Theta}_{ij}$ represents the angular position of cell
$ij$.

The advantage of this approach is obvious: instead of considering
$N^2$ pairs ($N$ is the number of galaxies) we have to consider only
$N_c^2$ pairs, where $N_c$ is the number of grid cells. Thus, the
number of pairs depends only on the cell size and not on the number of
galaxies. Hence, this method pays off once the cell size becomes large
enough, making the number of cells smaller than the number of
galaxies.  Moreover, in order to find all galaxies at some distance
from a certain cell we no longer have to check all galaxies, but only
neighbouring cells which are easy to find by the grid index.

The approach becomes inaccurate, however, for small angular bins,
because for these the assumption that cell-galaxies are essentially
concentrated into one single point is particularly inaccurate. By
comparing the ensemble average of $\hat{\xi}_+$ with the theoretical
$\xi_+$ we find that after the third angular bin this approximation
becomes accurate enough.  For our purposes, this approach is
completely sufficient. A better and more sophisticated approach can be
found in Pen~\&~Zhang (2003).

\begin{figure}[width=88mm]
  \epsfig{file=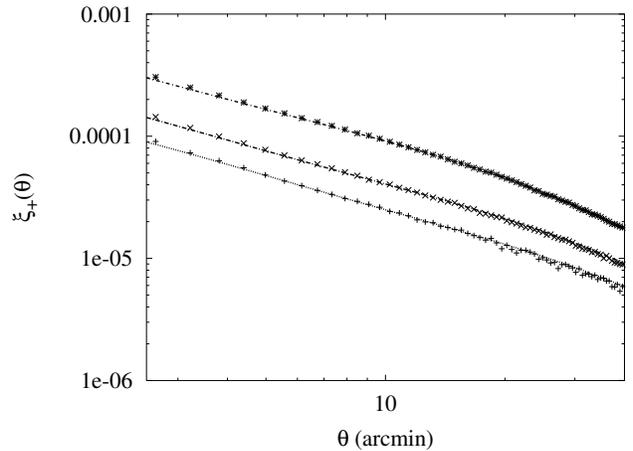, width=88mm}
\caption{Comparison of the analytical (lines)
  and numerical (symbols) shear correlation and cross-correlation
  functions $\xi_{\rm L, LU, U}$ (lower, middle and upper
  lines/symbols).}
\label{corr}
\end{figure}

For the case with $N_{z}=2$, with the division at $z_{\rm cut}=1.25$
(2binsIV) Fig.\,\ref{corr} shows the close agreement between the
correlation and cross-correlation functions, averaged over 7950
sub-fields, with the analytical prediction for the fiducial
$\Lambda$CDM cosmological model, obtained from Eq.\,\Ref{corre}. Shown
are comparisons for the lower (L) and upper (U) redshift bins, and
cross-correlation (LU).  To account for finite field size in our
numerical work, $\ell_{\rm min}=2\pi/14.9^\circ$ in the integration.
As noted above, since there is no well-defined cut-off, this value of
$\ell_{\rm min}$ was determined by allowing it to vary while
performing a least-squares fit of $\xi^{\rm t}_{{\rm L, U, LU}}$ to
$\ave{\hat{{\xi}}_{{\rm L, U, LU}}}$, so obtaining the inverse
variance weighted mean $\ell_{\rm min}$.  A cut-off at high $\ell$ is
not critical since in this regime the power-spectrum amplitude is much
lower.

\begin{figure}[width=88mm]
  \epsfig{file=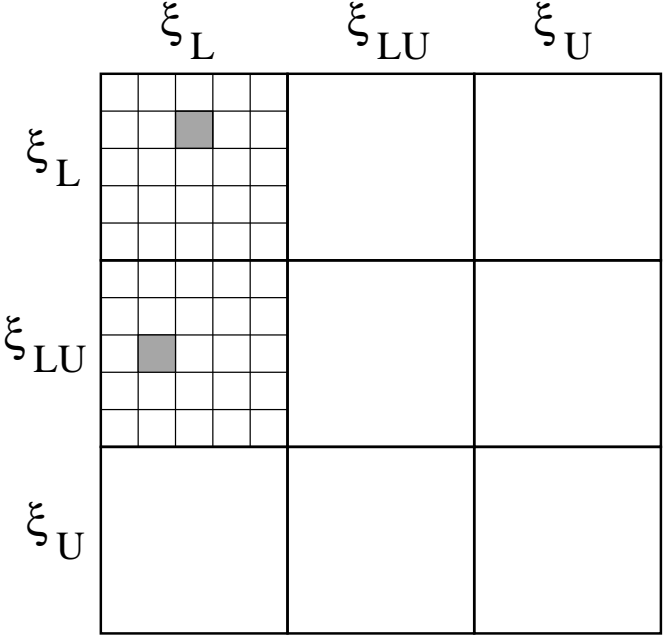, width=88mm}
\caption{Schematic illustration of the symmetric covariance matrix ${\bf C}$ for the case where there are $N_{z}=2$ source redshift bins and $N_{\Delta\theta}=5$ angular separation bins. Combinations of $\xi_{\rm L, U, LU}$ identify covariance terms of the form given in Eq.\,\Ref{coveq}.}
\label{cov}
\end{figure}

\begin{figure}[width=88mm]
  \epsfig{file=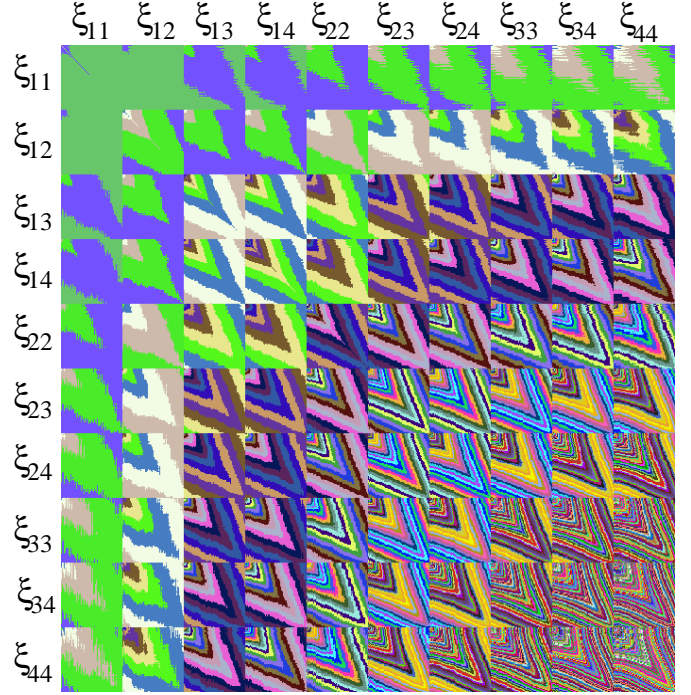, width=88mm}
\caption{The covariance matrix ${\bf C}$ determined from
  our simulations, for $N_{z}=4$, $N_{\Delta\theta}=65$ and a survey
  consisting of 1 sub-field. Different blocks correspond to auto- and
  cross-correlations between the redshift bins. Inside these blocks
  are auto-and cross-correlations for angular separation bins.}
\label{covey}
\end{figure}

\section{Estimating the covariance of $\hat{\xi}_\pm$}

We now outline how the covariance matrix of $\hat{\xi}_\pm$ is
estimated, for the case of $N_{z}=2$ redshift bins, with the division
at $z_{\rm cut}=1.25$ (2binsIV). As described above, our mock survey
consists of 10 uncorrelated fields. An angle bracket denotes averaging
over all 7950 sub-fields. Note that we may drop the L, U and LU
sub-scripts for ease of notation.

When no redshift binning is assumed, it is computationally
advantageous to determine the shear correlation function by combining
those determined for the case with redshift binning: \be
\hat{\xi}=n_{\rm L}^{2}\hat{\xi}_{{\rm L}}+2n_{\rm L}n_{\rm
  U}\hat{\xi}_{{\rm LU}}+n_{\rm U}^{2}{\hat{\xi}_{\rm U}}\,,
\label{xicompress}\ee where $n_{\rm L, U}$ are the fraction of sources
in the lower and upper bins respectively. A more general relation
between $\hat{\xi}$ and the cross- and auto-correlations of the shear
from more than two redshift bins can be found in appendix
\ref{xireduction}.

The impact of cosmic variance is taken into account by determining
correlation functions for each of the $N_{f}=795$ independent surveys.
The covariance matrix between bin $i$ and $j$ is determined using \be
{\bf C}_{ij}= \ave { \left( \hat{\xi}- \ave{\hat{\xi}} \right)_{i}
  \left( \hat{\xi}- \ave{\hat{\xi}} \right)_{j}}_{N_{f}} \;,
\label{coveq}
\ee where the outer average is performed over the $N_{f}=795$ surveys.
If redshift binning is considered, there are
$N_{z}\left(N_{z}+1\right)/2$ combinations of correlation and
cross-correlation functions and hence ${\bf C}$ is a matrix composed
of $\left[N_{z}\left(N_{z}+1\right)/2\right]^{2}$ blocks.
Fig.\,\ref{cov} illustrates this for the simplified case where
$N_{z}=2$ and $N_{\Delta\theta}=5$; for example the block in the upper
left of the matrix corresponds to elements $ {\bf C}_{ij}=\ave{ \left(
    \hat{\xi}_{\rm L}- \ave{\hat{\xi}}_{\rm L} \right)_{i}
  \left(\hat{\xi}_{\rm L}-\ave{\hat{\xi}}_{\rm L}\right)_{j} }$, and
the shaded entry to ${\bf C}_{2,3}$. The block in the middle row, left
column, corresponds to covariance elements between the
cross-correlation and the lower redshift bin, with the shaded entry
being ${\bf C}_{8,2}$. The bins denoted by $i$ and $j$ extend over
$\Delta\theta$ bins, repeated for each redshift auto- and
cross-correlation bin.

A representation of the covariance matrix determined from our
simulations with $N_{z}=4$ is shown in Fig.\,\ref{covey}.  Note that
${\bf C}$ has a strong diagonal, although it is not strictly
diagonally dominant.

Our covariance matrix for the case of no redshift binning is
consistent ($<10\%$ difference) with the treatment of Schneider et al.
(2002a), and with Kilbinger~\&~Schneider (2004) who adopted the same
assumption of Gaussianity.

\section{An application: Constraints on cosmological parameters}

The rather featureless two-point shear correlation function
$\xi_{+}(\theta)$ or corresponding convergence (shear) power spectrum
$P_{\kappa}(\ell)$ leads to strong degeneracies amongst the parameters
that can be derived from cosmic shear surveys.  An indication of the
degree of degeneracy is the behaviour of the partial derivatives of
$\xi_{+}$ with respect to each parameter $\pi_{i}$ (see King \&
Schneider 2003 for such a comparison), or using a Fisher matrix
analysis as in Sect.\,\ref{form}.

External sources of information often provide complimentary
constraints: for example, confidence regions in the $\Omega_{\rm
  m}-\sigma_{8}$ plane derived from weak lensing are almost orthogonal
to those from the analysis of CMB data (e.g. van Waerbeke et al.
2002), lifting this well known degeneracy (e.g. Bernardeau, van
Waerbeke \& Mellier 1997).

In this section we consider the extent to which crude redshift
information for sources used in a lensing analysis decreases the
expected errors in the $\Omega_{\rm m}-\sigma_{8}$, $\Omega_{\rm
  m}-\Gamma$ and $\sigma_{8}-\Gamma$ planes. Since we are interested
in the influence of redshift binning on parameter degeneracies, hidden
parameters are assumed to be perfectly known. As described above, we
focus on the information provided by the shear two-point correlation
function $\xi_{+}$ (we may drop the ``+'' subscript). We restrict this
application to the case of $N_{z}=2$ (2binsIV).  A larger parameter
space is then explored using the covariance matrix derived from
simulations in a Fisher analysis for the cases $N_{z}=2, 3, 4$.

\subsection{Obtaining confidence regions in the $\Omega_{\rm
    m}-\sigma_{8}$, $\Omega_{\rm m}-\Gamma$ and $\sigma_{8}-\Gamma$
  planes}

We now determine and compare the likelihood contours in the
$\Omega_{\rm m}-\sigma_{8}$, $\Omega_{\rm m}-\Gamma$ and
$\sigma_{8}-\Gamma$ planes for the cases with and without redshift
binning. The likelihood function is given by
\begin{eqnarray}\label{lik}
{\cal L}(\pi)&=&\frac{1}{(2\pi)^{n/2}|{\bf C}|^{1/2}}
\\
\nonumber&\times&\prod_{ij}{{\rm exp}\left[-\frac{1}{2}
\left(\xi^{\rm t}-\xi(\pi)\right)_{i}\left[{\bf C}^{-1}\right]_{ij}
\left(\xi^{\rm t}-\xi(\pi)\right)_{j}\right]}\;,
\end{eqnarray}
where $n$ is the number of rows (or columns) of the covariance matrix
${\bf C}$ and $\xi(\pi)$ are theoretical correlation functions
determined on a grid in parameter space.

\begin{figure}[width=88mm]
  \includegraphics[width=88mm]{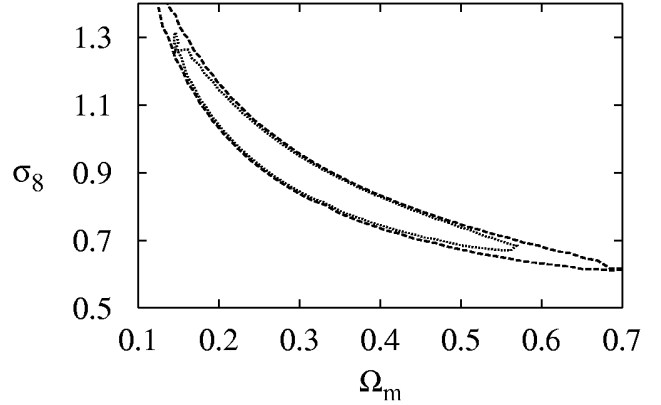}
\caption{Expected constraints in the $\Omega_{\rm m}-\sigma_{8}$ plane plotted
  for $\Delta\chi^{2}=4.61$ (90\% confidence) with (inner contour) and
  without (outer contour) redshift binning. The survey consists of 10
  square uncorrelated sub-fields, each $1.25^{\circ}$ on a side. The
  redshift distribution and binning are described in the text.}
\label{oms8p}
\end{figure}

\begin{figure}[width=88mm]
  \includegraphics[width=88mm]{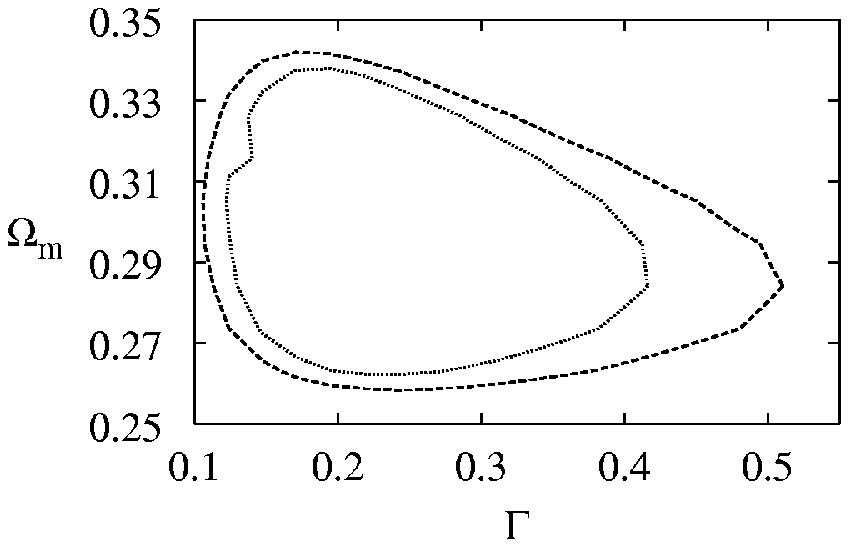}
\caption{Expected constraints in the $\Gamma-\Omega_{\rm m}$ plane plotted
  for $\Delta\chi^{2}=4.61$ (90\% confidence) with (inner contour) and
  without (outer contour) redshift binning. The survey is the same as
  in Fig.\,\ref{oms8p}.}
\label{omgap}
\end{figure}

\begin{figure}[width=88mm]
  \includegraphics[width=88mm]{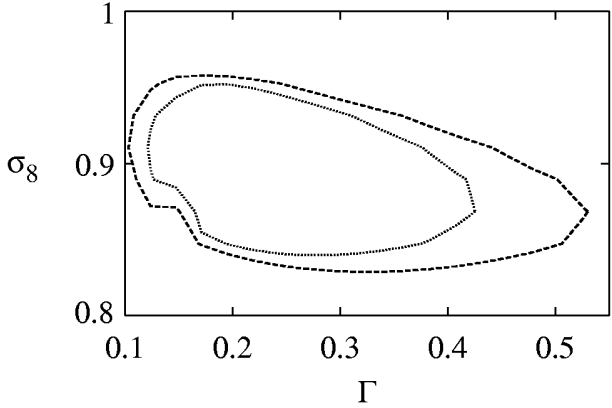}
\caption{Expected constraints in the $\Gamma-\sigma_{8}$ plane plotted
  for $\Delta\chi^{2}=4.61$ (90\% confidence) with (inner contour) and
  without (outer contour) redshift binning. The survey is the same as
  in Fig.\,\ref{oms8p}.}
\label{gas8p}
\end{figure}

The log-likelihood function is distributed as $\chi^{2}/2$ so that \be
\chi^{2}(\pi)=\sum_{ij}{\left(\xi^{\rm
      t}-\xi(\pi)\right)_{i}\left[{\bf C}^{-1}\right]_{ij}
  \left(\xi^{\rm t}-\xi(\pi)\right)_{j}}\;. \ee Confidence contours
can be drawn in this $\chi^{2}$-surface, relative to the minimum
(zero) at $\xi(\pi)\equiv\xi^{\rm t}$. In
Figs.\,\ref{oms8p}-\ref{gas8p} the confidence contours are shown for
each of the $\Omega_{\rm m}-\sigma_{8}$, $\Omega_{\rm m}-\Gamma$ and
$\Gamma-\sigma_{8}$ planes, with and without redshift binning. Note
that while $\Omega_{\rm m}$ is varied, we keep $\Omega_{\rm
  m}+\Omega_{\Lambda}=1$. To highlight the difference and avoid
confusion, we plot contours for a single value of $\Delta\chi^{2}$.


\subsection{Fisher information}

The Fisher matrix (Fisher 1935) gives a handle on the question as to
how accurately model parameters can be estimated from a given data
set. In this section, we will use this method to examine
quantitatively the increase of information on the cosmological
parameters $\Omega_{\rm m}$, $\Omega_\Lambda$, $\sigma_8$ and $\Gamma$
when the number of redshift bins, thus the knowledge of the
three-dimensional distribution of the galaxies, is increased. Note
that we no longer impose the condition $\Omega_{\rm
  m}+\Omega_\Lambda=1$. After a brief introduction to this topic we
apply the Fisher statistics to our simulated data.

\subsubsection{Fisher Formalism\label{form}}

In general, one uses data points $\xi_i$ from a measurement
to infer model parameters $\pi_i$ based on a theoretical model. As
the measurements are polluted by noise, we cannot expect to exactly
obtain the data points $\xi\left(\pi\right)$ predicted by our model.
But we can try to find a combination of model parameters
$\hat{\pi}_i$ that predict data points as close as possible to the
actual measurement.  The closeness is decided on the grounds of a
statistical estimator.  The covariance of the parameter
uncertainties
\begin{equation}
  Q_{ij}\equiv\ave{\Delta\pi_i\Delta\pi_j}
\end{equation}
with
$\Delta\pi_i\equiv\left(\ave{{\hat\pi}_i^2}-\ave{{\hat\pi}_i}^2\right)^{1/2}$
is related to the so-called Fisher information matrix through
\begin{equation} \label{fisher}
  F_{ij}\equiv
  -\ave{\frac{\partial^2\log{\cal
        L\left[\pi,\xi\right]}}
    {\partial\pi_i\partial\pi_j}}
  = \ave{\left[\Matrix{Q}^{-1}\right]_{ij}}
  \; .
\end{equation}
$\cal L$ corresponds to the likelihood for obtaining the measurement
$\xi$ keeping the underlying model parameter $\pi_i$ fixed. See for
example Tegmark et al. (1997) and references therein for a more
detailed description.

It follows from statistics that the $1\sigma$ scatter of the estimated
parameters is (Cram\'{e}r-Rao inequality)
\begin{equation}
  \Delta\pi_i\ge\sqrt{\left[\Matrix{F}^{-1}\right]_{ii}}
  \; ,
\end{equation}
where commonly the lower limit is taken to be the estimate for
$\Delta\pi_i$. To quantify the degeneracies in the parameter estimate,
we evaluate the correlation of the estimate's uncertainty contained in
$\Matrix{F}$:
\begin{equation}
  r_{ij}\equiv
  \frac{\ave{\Delta\pi_i\Delta\pi_j}}
  {\sqrt{\ave{\Delta\pi_i^2}\ave{\Delta\pi_j^2}}}=
  \frac{\left[\Matrix{F}^{-1}\right]_{ij}}
  {\sqrt{\left[\Matrix{F}^{-1}\right]_{ii}\left[\Matrix{F}^{-1}\right]_{jj}}}
\end{equation}
as, for example, in Huterer (2002).  Highly correlated or
anti-correlated $\Delta\pi_i$ and $\Delta\pi_j$ are called degenerate,
whereas no correlation means no degeneracy (for the fiducial model).
Another piece of information that can be extracted from the Fisher
matrix is the orientation of the error ellipsoid in parameter space,
which is defined by the eigenvectors of $\Matrix{F}$.  This
corresponds to the directions of degeneracies.

In the application of this formalism in Sect.\,\ref{app} below, we
will look at situations where some of the model parameters are assumed
to be known a priori. In this case, they are no longer free parameters
that have to be estimated from measured data points, so that the size
of the Fisher matrix reduces according to the number of parameters
fixed.  This amounts to removing rows and columns from the general
Fisher matrix, one for each fixed parameter, so that these cases can
be considered by simply looking at sub-matrices of the largest Fisher
matrix. Taking all conceivable sub-matrices enables the exploration of
all possible combinations of fixed (strong prior) and free parameters.

In practice, one uses the approximation \Ref{lik} for the likelihood
function $\cal L$ (where $\xi^{\rm t}\equiv\xi$), so that the Fisher
information matrix is approximately
\begin{equation}
  F_{ij}=
  \sum_{kl}
  \left[\frac{\partial\xi\left(\pi\right)}{\partial\pi_i}\right]_k
  \left[\Matrix{C}^{-1}\right]_{kl}
  \left[\frac{\partial\xi\left(\pi\right)}{\partial\pi_j}\right]_l
  \; ,
\label{fish}
\end{equation}
which is exact in the case of pure Gaussian statistics, but
  may be used as a good approximation for the valley in parameter
  space in which the minimum of $-\log{\cal L}$ lies.  Again,
$\Matrix{C}$ is the covariance of the measured data points and
$\xi\left(\pi\right)$ the vector of modelled data points in absence of
noise.

\subsubsection{Application of the Fisher formalism\label{app}}
Now we use the Fisher formalism to estimate constraints on various
combinations of parameters, with different degrees of redshift
binning.  First, we evaluate Eq.\,\Ref{fish} using the covariance
matrix from our fiducial survey consisting of 10 independent
sub-fields, $N_{z}=2$ (with $z_{\rm cut}=1.25$), and
$N_{\Delta\theta}=65$. The procedure is repeated for the covariance
matrix for the coarser $N_{z}=1$ binning.  Table\,\ref{table1} shows
the percentage error for $N_{z}=2$ as opposed to $N_{z}=1$ for the
same set of free and fixed parameters.

We extend the treatment to a larger number of redshift bins, in the
context of the survey consisting of 1 sub-field. Again Eq.\,\Ref{fish}
is calculated, this time using the covariance matrices from the
simulations with $N_{z}=4$, and those from coarser binning ($N_{z}=3,
2, 1$) of this data set. Table\,\ref{table2} lists the errors for
$N_{z}=4, 3, 2$ as a percentage of the $N_{z}=1$ error.

\begin{center}
\begin{figure}[width=75mm]
  \epsfig{file=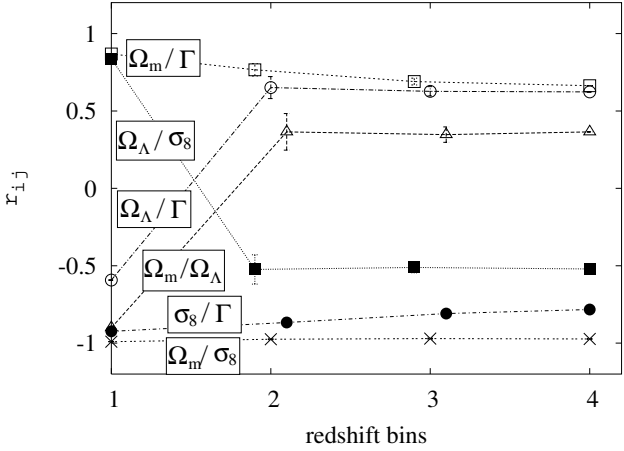,width=90mm,angle=0}
\caption{Correlations of the errors in the parameter estimates for
  different pairs of parameters and numbers of redshift bins as
  derived from the Fisher matrix; only the case with no fixed priors
  is considered. Error bars denote the variance in the correlations
  for the different redshift binnings for the same number of bins
  (only for 2 and 3 bins). The data points are slightly shifted to
  avoid overlapping.}
\label{degeneracy}
\end{figure}
\end{center}

In order to investigate the degeneracies of the parameter estimates,
we concentrate on the case that no priors are given. For this
particular situation, the gain by introducing redshift binning is
largest (see Tables\,\ref{table1} and \ref{table2}). In
Fig.\,\ref{degeneracy} we plot the correlations of the errors in the
parameter estimates for different pairs of parameters and different
numbers of redshift bins. If more than one redshift binning for the
same number of redshift bins is available in our data set, we indicate
the scatter of correlation coefficients by error bars.  Some scatter
indicates that the correlations can be changed slightly by varying the
bin limits. The strong correlation between the estimates of
$\Omega_{\rm m}$ and $\sigma_8$ is only marginally affected by
redshift binning.  This is also the case for fixed $\Gamma$ and/or
$\Omega_\Lambda$ (not shown).

 \begin{table}
  \begin{center}
 \caption{Uncertainties in the parameter estimates according to the Fisher
   formalism, for our fiducial survey of 10 uncorrelated sub-fields.
   The first data set (2binsIV) with $N_{z}=2$ and $z_{\rm cut}=1.25$
   is used. Columns with dots ``.'' denote fixed parameters (strong
   priors).  Uncertainties in the top panel are absolute values for a
   single redshift bin. Those in the lower panel are for $N_{z}=2$,
   quoted as a percentage of the single redshift bin ($N_{z}=1$) case.
   For instance, with no fixed parameters, $\Delta\Omega_\Lambda=0.26$
   with $N_{z}=2$ (i.e. 13\% of the $N_{z}=1$ error).
   $|\Matrix{F}^{-1}|$ denotes the determinant of the inverse of the
   Fisher matrix; its square root is proportional to the volume of the
   error ellipsoid in parameter space. The $n$th root, with $n$ being
   the number of free parameters, defines a typical size of the error
   ellipsoid; this size is proportional to the geometric mean of the
   lengths of the principal ellipsoid axes.
\label{table1}}
  \begin{tabular}{ccccc} 
    $\Delta\Omega_m$ & $\Delta\Omega_\Lambda$ & $\Delta\sigma_8$ &
    $\Delta\Gamma$ & $\sqrt{|\Matrix{F}^{-1}|}^{1/n}$ 
    \\
    \hline
    \\
    0.9 & 2.0 & 1.2 & 0.4    & 0.16\\
    .      & 0.5 & 0.1 & 0.1 & 0.09 \\
    0.2 & .      & 0.4 & 0.2 & 0.04 \\
    0.09 & 0.7 & .      & 0.08 & 0.08 \\
    0.3 & 1.0 & 0.3 & .      & 0.13 \\
    \\
    0.06 & 0.5 & .      & .       & 0.09 \\
    0.08 & .      & 0.1 & .        & 0.05 \\
    0.02 & .      & .      & 0.07 & 0.03 \\
    .      & 0.3 & 0.06 &   .     & 0.09 \\
    .      & 0.1 & .      & 0.06  & 0.08 \\
    .      &      . & 0.03 & 0.07 & 0.04 \\
    \\
    0.02 & .      &  .     & .      & 0.15 \\
    .      & 0.10  & .      &   .    & 0.10  \\
    .      &      . & 0.02 & .      & 0.02 \\
    .      &      . &      . & 0.06 & 0.06  \\
    \hline
    \\
    13\% & 13\% & 17\% & 26\% & 47\% \\
    .      & 52\% & 57\% & 64\% & 72\% \\
    53\% & .      & 52\% & 49\% & 73\% \\
    43\% & 38\% & .      & 68\% & 65\% \\
    32\% & 24\% & 46\% & .      & 57\% \\
    \\
    50\% & 45\% & .      & .      & 64\% \\
    86\% & .      & 85\% & .      & 88\% \\
    89\% & .      & .      & 81\% & 85\% \\
    .      & 65\% & 71\% &   .    & 76\% \\
    .      & 80\% & .      & 81\% & 81\% \\
    .      &      . & 87\% & 80\% & 84\% \\
    \\
    85\% & .      &  .     & .      & 85\% \\
    .      & 81\% & .      &   .    & 81\% \\
    .      &      . & 89\% & .      & 89\% \\
    .      &      . &      . & 82\% & 82\%
  \end{tabular}
  \end{center}
 \end{table}
\begin{table}
  \begin{center}
  \caption{The notation is identical to Table \ref{table1}. The second
    data set was used to obtain these values, with various binnings in
    redshift denoted by the final column, and with covariance matrices
    calculated for 1 sub-field. Uncertainties are again quoted as a
    percentage of that for $N_{z}=1$.\label{table2}}
  \begin{tabular}{cccccr}
    $\Delta\Omega_m$ & $\Delta\Omega_\Lambda$ & $\Delta\sigma_8$ &
    $\Delta\Gamma$ & $\sqrt{|\Matrix{F}^{-1}|}^{1/n}$ & data set
    \\
    \hline
    \\
    16\%  & 14\% & 19\% & 26\%  & 41\% & 2binsI \\
    17\%  & 14\% & 21\% & 32\%  & 42\% & 2binsII \\
    21\%  & 18\% & 30\% & 53\% & 49\% & 2binsIII \\
    \\
    85\% & . & 84\% & . & 88\% & 2binsI \\
    80\% & . & 80\% & . & 85\% & 2binsII \\
    88\% & . & 89\% & . & 91\% & 2binsIII \\
    \\
    74\% & . & . & 70\% & 79\% & 2binsI \\
    75\% & . & . & 74\% & 82\% & 2binsII \\
    92\% & . & . & 91\% & 92\% & 2binsIII \\
    \\
    . & . & 67\% & 63\% & 76\% & 2binsI \\
    . & . & 71\% & 69\% & 79\% & 2binsII \\
    . & . & 90\% & 87\% & 91\% & 2binsIII \\
    \hline
    13\% & 11\% & 16\% & 23\% & 34\% & 3binsI \\
    13\% & 12\% & 16\% & 22\% & 36\% & 3binsII \\
    14\% & 11\% & 18\% & 27\% & 36\% & 3binsIII \\
    \\
    73\% & . & 75\% & . & 79\% & 3binsI \\
    73\% & . & 73\% & . & 79\% & 3binsII \\
    72\% & . & 73\% & . & 78\% & 3binsIII \\
     \\
    68\% & . & . & 65\% & 73\% & 3binsI \\
    67\% & . & . & 64\% & 74\% & 3binsII \\
    69\% & . & . & 67\% & 74\% & 3binsIII \\
     \\
    . & . & 62\% & 58\% & 70\% & 3binsI \\
    . & . & 61\% & 57\% & 70\% & 3binsII \\
    . & . & 65\% & 61\% & 72\% & 3binsIII \\
    \hline
    11\% & 9\% & 14\% & 19\% & 31\% & 4bins \\
    65\% & . & 67\% & . & 71\% & 4bins \\
    61\% & . & . & 57\% & 66\% & 4bins \\
    . & . & 55\% & 51\% & 63\% & 4bins \\
  \end{tabular}
  \end{center}
\end{table}
\section{Discussion}
The average shear correlation functions obtained from our numerical
simulations are in good agreement with those obtained analytically, as
was illustrated in Fig.\,\ref{corr}. We also pointed out that their
covariance is compatible with Schneider et al. (2002a) and
Kilbinger~\&~Schneider (2003).

Our treatment is only strictly valid for Gaussian density
fields and is a good approximation for scales greater than $\sim
10\arcmin$, giving a lower limit on the covariance at smaller
scales (e.g. van Waerbeke et al. 2002).  A more accurate covariance
matrix is possible, though. According to Schneider et al. (2002a)
(section 4 therein), the covariance matrix of $\xi_+$ may be
decomposed into three terms
\begin{equation}
  C_{ij}=\sigma_\epsilon^2X_{ij}+\sigma_\epsilon^4Y_{ij}+Z_{ij}
  \; ,
\end{equation}
where $X$, $Y$ and $Z$ are some functions. $X$ and $Y$ are functions
of the two-point correlation of cosmic shear and consequently
insensitive to non-Gaussian features of the field. $Z$, however,
depends linearly on the four-point correlation of cosmic shear which
in Schneider et al. (2002a) is worked out by assuming a Gaussian
field; this factorises $Z$ into a sum of products of two-point
correlators only. In the hierarchical clustering regime, the
four-point correlation of the random field differs from that value
only by a constant scale-independent factor $Q$, the so-called
hierarchical amplitude (see e.g. Bernardeau et al. 2002). Thus,
hierarchical clustering increases the component $Z$ simply by the
factor $Q$. This could be included in $C_{ij}$ by the following two
steps: i) calculating $C_{ij}\equiv C^{\left(1\right)}_{ij}$ by setting the
intrinsic noise $\sigma_\epsilon=0$, ii) recalculating
$C_{ij}\equiv C^{\left(2\right)}_{ij}$ this time with the intrinsic noise turned
on. The final covariance matrix $C_{ij}$, accounting for $Q$, is
obtained by
\begin{equation}
  C_{ij}=C^{\left(1\right)}_{ij}\left(Q-1\right)+C^{\left(2\right)}_{ij}
  \; .
\end{equation}

As an illustration of the use of numerically derived covariance
matrices, we have considered to what degree redshift information
tightens the confidence regions on cosmological parameters.  Note that
we required rather crude redshift information: photometric redshift
estimates for sources can be obtained from multi-colour observations,
with typical accuracies in $\sigma_{z}$ of $\sim 0.1$ or better (e.g.
Bolzonella et al. 2000).

We have investigated the information contained in $\xi_{+}$, rather
than the shear (or equivalently convergence) power spectrum
$P_{\kappa}$, since it is directly obtained from the statistics of the
distorted images of distant galaxies. Various estimators of
$P_{\kappa}$ have been proposed, requiring the {\em spatial}
distribution of the shear (e.g. Kaiser 1998; Hu \& White 2001) or the
shear correlation functions (Schneider et al. 2002a). Note that a
calculation similar to ours but using ${\hat P}_{\kappa}$ would
formally require one to use the full covariance matrix for the
associated estimator. However, as noted in Schneider et al. (2002a),
the band-power estimates for ${\hat P}_{\kappa}$ do decorrelate rather
quickly.

The first data set consists of 2 redshift bins, and covariance
matrices estimated with and without binning for a survey with 10
uncorrelated (1.25$^{\circ}$ on a side) sub-fields (i.e. selected from
different realisations).  Constraints on pairs of cosmological
parameters using a likelihood treatment, with and without redshift
information, are shown in Figs.\,\ref{oms8p}-\ref{gas8p}. Since our
goal here is to study the benefit of redshift information in cosmic
shear studies, we do not adopt priors from WMAP or other probes of
large scale structure which might confuse the issue. In both cases the
redshift {\em distribution} is assumed to be known. Assigning sources
to 2 redshift bins tightens the confidence regions in all cases.

Noteworthy are the tightened upper limits on $\Gamma$ in the
$\Gamma-\sigma_{8}$ and $\Omega_{\rm m}-\Gamma$ planes when binning is
employed.  The constraints on $\Gamma$ in both planes are rather
asymmetric, with the confidence regions being more extended towards
high $\Gamma$ values. $\Gamma$ determines the location of the peak in
the matter power spectrum, and having extra redshift information
places tighter constraints on this - there is a degeneracy between
$\Gamma$ and the mean source redshift $\ave{z}$ such that a larger
$\Gamma$ would be compensated for by a smaller $\ave{z}$.  Recall that
$\Gamma$ is not a fundamental quantity; in the limit of zero baryons,
$\Gamma=\Omega_{\rm m}h$, which is modified to $\Gamma=\Omega_{\rm
  m}\,h\,{\rm exp}\left(-\Omega_{\rm b}\left(1+\sqrt{2h}/\Omega_{\rm
      m}\right)\right)$, if $\Omega_{\rm b}$, the present baryon
density, is accounted for in the transfer function (Sugiyama 1995).
With our strong priors (hence fairly tight constraint on $\Omega_{\rm
  m}$ or $\sigma_{8}$), the error in $\Gamma$ roughly translates into
an error in $h$, so redshift binning decreases the upper limit on $h$.

In the $\Omega_{\rm m}-\sigma_{8}$ plane we obtain the familiar
``banana'' shaped confidence regions, tightened with the inclusion of
redshift binning. It is difficult to directly compare our constraints
to real surveys with different observational conditions; however, our
confidence regions are roughly compatible with those of van Waerbeke
et al. (2001) allowing for these differences.

To explore a wider range of parameter combinations, we employed the
Fisher formalism since this allows one to easily obtain error
estimates and investigate degeneracies. We used the covariance
matrices estimated from the first data set, again for 10 sub-fields as
described above.  Table \ref{table1} shows to what extent the errors
on various parameters are improved for $N_{z}=2$ compared with
$N_{z}=1$. Note that these results depend on the cosmological model
and power spectrum of our fiducial model.  Redshift binning is
particularly helpful when fewer strong priors are assumed, compared
with the case when only one or two of the parameters are allowed to
vary. In the case where $\Omega_{\rm m}$, $\Omega_{\Lambda}$,
$\sigma_{8}$ and $\Gamma$ are free, errors are a factor of roughly 4
to 8 smaller when $N_{z}=2$.  As we adopt more strong priors, redshift
binning becomes progressively less beneficial. For example, if we
consider parameter combinations where either $\Omega_{\rm m}$ or
$\sigma_{8}$ are assumed to be perfectly known, this breaks a strong
degeneracy otherwise present; the decrease in errors when $N_{z}=2$
are therefore not so great as one might have anticipated. Another
interesting trend is that the constraint of $\Omega_{\Lambda}$ seems
to be most favourably affected by redshift binning, perhaps because it
is important to the growth rate of structure at redshifts $z\sim 1$.

How does Fisher analysis compare with the likelihood treatment? Fisher
analysis should be seen as a way to estimate errors and investigate
degeneracies, but does not propose to reveal the detailed behaviour of
confidence regions far from the fiducial model -- for instance the
asymmetric constraints in the $\Gamma-\Omega_{\rm m}$ or $-\sigma_{8}$
planes, or the ``banana'' shaped constraints in the $\Omega_{\rm
  m}-\sigma_{8}$ plane. Nevertheless, the differences in the
constraints on the parameter pairs drawn from $\Omega_{\rm m}$,
$\sigma_{8}$ and $\Gamma$ are roughly consistent with the full
likelihood treatment discussed above.

Using the second data set consisting of 4 redshift bins, for a survey
consisting of 1 sub-field, we obtained covariance matrices
corresponding to various redshift cuts for 2 and 3 redshift bins.
Parameter constraints are again tightened going from $N_{z}=1$ to
$N_{z}=4$. Considering again for example the case where all parameters
are free, the most dramatic change is already seen going from
$N_{z}=1$ to $N_{z}=2$. In the case of $N_{z}=2$, with the lowest
redshift cut $z_{\rm cut}=0.75$ (2binsIII), parameter constraints are
weakest. When higher redshift sources are isolated ($z_{\rm cut}=1.5,
2.25$) the constraints are similar, and which is the better choice
depends on the combination of parameters considered. For 3 redshift
bins, the combinations of cuts in redshift lead to very similar error
estimates.

At some point further sub-division into redshift bins does not lead to
improved constraints on parameters (Hu 1999). This limit must be
determined for the survey and cosmological parameters in question.
For simplicity consider the case of $N_{z}=2$: $\xi_{\rm L}$ and
$\xi_{\rm U}$ are correlated since sources in the upper redshift bin
are also sensitive to lensing by structure at $z<z_{\rm cut}$.  A
measure of this correlation is provided by a correlation parameter,
$p=\xi_{\rm LU}/\left(\xi_{\rm L}\xi_{\rm U}\right)^{0.5}$, also used
by Hu (1999) in the context of power spectra.  $p=1$ and $p=0$
indicate complete correlation and lack of correlation respectively.
For our simulations, $0.83<p<0.93$ for the range over which the
correlation functions are calculated, with $\bar{p}=0.87$ taking the
mean over all $\Delta\theta$ bins.

The improvement of the parameter constraints is clearly also a
function of how the redshift intervals are set. This can already be
seen in Table \ref{table2} where we calculated different binnings
with two and three redshift bins. We expect that there is an optimal
way to bin the data. The intention of this paper, however, is to
present a fast method for calculating the covariance matrix of the
cosmic shear correlation estimator. This issue will be explored in a
forthcoming publication. In practise, this question probably does
not arise anyway, because there one would take as many redshift bins
as possible, their number being determined by the 
accuracy of the redshift estimator.

As described above, one of our fiducial surveys is composed of 10
uncorrelated sub-fields. Another possibility is that a single
contiguous patch of sky is targeted primarily for another science
goal; such a survey might consist of 10 sub-fields drawn from the same
large field. The sub-fields might in that case be selected so as to
avoid bright stars, chip boundaries or defects. These sub-fields would
be correlated to some extent, meaning that taking $m$ rather than
$n(<m)$ sub-fields does not decrease the covariance by a factor of
$m/n$, which would be the case if they were independent. The degree to
which the sub-fields are correlated is accounted for directly in the
covariance matrix. However, it is instructive to have an estimate of
this, using the (ensemble average) covariance matrix for a survey
composed of 10 {\em correlated} as opposed to 10 {\em uncorrelated}
sub-fields. The ratio of the diagonal elements of the correlated and
uncorrelated geometry covariance matrices ranges between $\approx
1.1-1.4$, from the lower to upper redshift bin. Of course, the amount
of correlation between sub-fields also depends on their geometry, so
this has to be estimated for the survey in question.
\section{Conclusions}
In preparation for upcoming cosmic shear surveys, the main purpose of
this paper was to demonstrate that it is possible to rapidly simulate
surveys using a Monte Carlo method. This enables one to obtain
accurate {\em full} covariance matrices for the shear two-point
correlation function $\xi_{+}$, estimated from arbitrary survey
geometries and with the sources binned in redshift. Averaging over
many independent realisations enables us to take into account cosmic
variance.  As a first application, we estimated the extent to which
redshift information for sources used in a cosmic shear analysis
improves constraints on parameters derived from the estimated shear
two-point correlation function $\hat{\xi}_{+}$.
  
A likelihood analysis in the $\Omega_{\rm m}-\sigma_{8}$, $\Omega_{\rm
  m}-\Gamma$ and $\Gamma-\sigma_{8}$ planes shows that separating the
sources into two redshift bins enables tighter constraints to be
placed on parameters. Considering a wider range of parameter
combinations in the context of a Fisher analysis reveals that redshift
information is particularly advantageous in cases where few strong
priors are assumed.  When $\Omega_{\rm m}$, $\sigma_{8}$, $\Gamma$ and
$\Omega_{\Lambda}$ are free parameters, having 2 (4) redshift bins
tightens errors on parameters by a factor of $\sim 4-8$ ($\sim 5-10$).
Most improvement on error estimates occurs going from $N_{z}=1$ to
$N_{z}=2$.  In general, for the combinations of free and fixed
parameters explored, $\Omega_{\Lambda}$ seems to benefit most from
redshift binning,

One might ask why cosmic shear is of interest, in the light of the
recent WMAP results (e.g. Bennett et al. 2003), which suggest that the
era of precision cosmology is already upon us; there are several
facets to consider. Cosmic shear has the power to break degeneracies
inherent to CMB data (e.g. Hu \& Tegmark 1999), for instance the
angular diameter distance degeneracy (e.g. Efstathiou \& Bond 1999).
It also provides a completely independent cross-check of cosmological
parameters, based on equally well understood but different physical
principles. Besides this, as is the case with the CMB, the
interpretation of results requires no assumption about the bias
between luminous tracers and the underlying dark matter distribution
which plagues, for example, galaxy redshift surveys. In addition, with
redshift estimates for sources in upcoming large cosmic shear surveys,
lensing has the potential to see beyond the radially projected
convergence power spectrum, becoming sensitive to the evolution and
growth of structure in the Universe.

\begin{acknowledgements}
  We would like to thank Martin Kilbinger, Ludo van Waerbeke and Wayne
  Hu for helpful discussions. This work was supported by the Deutsche
  Forschungsgemeinschaft under the Graduiertenkolleg 787 and the
  project SCHN 342/3--1, and by the German Ministry for Science and
  Education (BMBF) through the DLR under the project 50 OR 0106.
\end{acknowledgements}

\appendix
\section{Switching from a finer to a coarser redshift binning\label{xireduction}}
Here we show how the auto- and cross-correlations of the cosmic shear
from a finer redshift binning are related to the auto- and
cross-correlations obtained from a coarser redshift binning (by
combining the finer bins). The relations of this section ensure that
we only have to make simulated data of the finest redshift binning,
since the corresponding data with less redshift information can always
be related to the 3D-correlations of the cosmic shear of this case.

In a first step, we turn to the auto-correlation $\xi_\pm$ of a new
redshift bin, neglecting for a moment the cross-correlations to the
other new redshift bins.  $\xi_\pm$ is according to the
Eqs.\,\Ref{kappow} and \Ref{corre} a function that linearly depends on
$\bar{W}^2\left(w\right)$
\begin{eqnarray}
\xi_\pm\left(\theta\right)&=&
\frac{9H_0^4\Omega_m^2}{4c^4}
\\\nonumber
&\times&\int\frac{\d\ell\ell}{2\pi}\int_0^{w_h}\d w
 \frac{\bar{W}^2\left(w\right)}{a^2\left(w\right)}
 J_{0,4}\left(\ell\theta\right)
 P_\delta\left(\frac{\ell}{f\left(w\right)},w\right)
 \; .
\end{eqnarray}
If we split the redshift distribution $p\left(w\right)$ of the source
galaxies into disjunct parts, like
\begin{eqnarray}\nonumber
  p\left(w\right)&=&\sum_i
  q^{\left(i\right)}\left(w\right)\\
  q^{\left(i\right)}\left(w\right)&\equiv&
  \left\{
    \begin{array}{ll}
      p\left(w\right)&~~{\rm for}~w\in\left[w_{i-1},w_i\right]\\
      0&~~{\rm else}
    \end{array}
  \right.
\end{eqnarray}
with $w_i(z_i)$ in the sense of Fig.\,\ref{redshiftbins}, we expand
with the notation of Eqs.\,\Ref{powerspectra} the function $\bar{W}$
in the following manner
\begin{eqnarray}\label{wdivide}
\bar{W}^{2}\!\!\!\!\!&&\!\!\!\!\!\left(w\right)=
\sum_{ij}n_in_j
\bar{W}^{\left(i\right)}\left(w\right)
\bar{W}^{\left(j\right)}\left(w\right)
\\\nonumber
&=&  \sum_i n_i^2
  \left[\bar{W}^{\left(i\right)}\left(w\right)\right]^2+
  2\sum_{i>j}n_i n_j
  \bar{W}^{\left(i\right)}\left(w\right)
  \bar{W}^{\left(j\right)}\left(w\right)
  \; .
\end{eqnarray}
As the $q^{\left(i\right)}$ defined here are not normalised, but the
corresponding redshift distributions $p^{\left(i\right)}$ in the
definition for $\bar{W}^{\left(i\right)}\left(w\right)$ are, we
introduce the \emph{normalisation factors}
\begin{equation}
  n_i\equiv\int_{w_{i-1}}^{w_i}\d w~
  q^{\left(i\right)}\left(w\right)
\end{equation}
telling us what fraction of the distribution inside the new bin is
contained in its subdivisions $q^{\left(i\right)}$.  Translating
Eq.\,\Ref{wdivide} to the power spectra gives, leaving out the
arguments in $\ell$
\begin{equation}\label{pdivide}
P_\kappa=
\sum_i n_i^2\,
P_\kappa^{\left(ii\right)}+
2\sum_{i>j}n_i n_j\, P_\kappa^{\left(ij\right)}
\; .
\end{equation}
Similarly, we get for the cosmic shear auto-correlation $\tilde{\xi}$
(actually for all linear functions of $P_\kappa^{\left(ij\right)}$;
hence the dropped index ``$\pm$''):
\begin{equation}\label{xidivide}
\tilde{\xi}=
\sum_i n_i^2\,
\xi_{ii}+
2\sum_{i>j}n_i n_j\, \xi_{ij}
\;,
\end{equation}
where $\xi_{ii}$ are the auto-correlations for the sub-bins,
$\xi_{ij}$ are the cross-correlations of the cosmic shear between the
sub-bins.  This equation tells us, therefore, how we have to combine
the cosmic shear correlations of the sub-bins when we are switching
from a finer to a coarser redshift binning of the data.

What about the cross-correlations between the new redshift bins if we
decide to switch to a binning with more than one redshift bin?  This
case is treated like the foregoing one, except that it is slightly
more general. Assume we focus on two new redshift bins $k$ and $l$
consisting of data from a finer redshift binning:
\begin{equation}
  p_k\left(w\right)=\sum_i
  q_k^{\left(i\right)}\left(w\right)
  \; ;\;\;
  p_l\left(w\right)=\sum_j
  q_l^{\left(j\right)}\left(w\right)
  \; ,
\end{equation}
where $p_k$ is the redshift distribution inside the new bin $k$ and
$p_l$ the same for the new bin $l$. Bin $k$ combines
$q_k^{\left(i\right)}$ and the bin $l$ combines $q_l^{\left(j\right)}$
from a finer binning, respectively. Using the same arguments as
before, we obtain the following relation between the cosmic shear
cross-correlation between the new redshift bins $\tilde{\xi}_{kl}$,
and the cross-correlations $\xi^{\left(kl\right)}_{ij}$ between their
components:
\begin{equation}\label{generalxidivide}
  \tilde{\xi}_{kl}=
  \sum_{ij}n^{\left(k\right)}_i n^{\left(l\right)}_j\xi^{\left(kl\right)}_{ij}
  \; .
\end{equation}
$n^{\left(k\right)}_i$ and $n^{\left(l\right)}_j$ are the
normalisations for the sub-bins. This equation is, of course, the
generalisation of Eq.\,\Ref{xidivide}.

\end{document}